\documentstyle[12pt]{article}
\def\q{$q$}
\def\qbar{$\bar{q}$}
\def\g{$g$}

\def\als{$\alpha_s$}
\def\d{{\rm d}}
\def\ds{\displaystyle}

\def\y0{$y_0$}
\def\yj{$y_J$}
\def\yg{$y_\gamma$}
\def\ms{$\overline{{\rm MS}}$}
\def\pQCD{perturbative QCD}

\def\np#1#2#3  {{\it Nucl. Phys. }{\bf #1} (19#3) #2}
\def\pl#1#2#3  {{\it Phys. Lett. }{\bf #1} (19#3) #2}
\def\pr#1#2#3  {{\it Phys. Rev. }{\bf #1} (19#3) #2}

\def\prl#1#2#3 {{\it Phys. Rev. Lett. }{\bf #1} (19#3) #2}
\def\rmp#1#2#3 {{\it Rev. Mod. Phys. }{\bf #1} (19#3) #2}
\def\zp#1#2#3  {{\it Zeit. Phys. }{\bf #1} (19#3) #2}

\begin{document}

\begin{titlepage}
\setcounter{footnote}{1}
\begingroup
\def\thefootnote{\fnsymbol{footnote}}
\vspace*{-2cm}
\begin{flushright}
ETH-TH/92-26\\
June 1992 \\
\end{flushright}
\vskip .5in
\begin{center}
{\Large\bf
QCD Corrections to  Photon Production in Association
with Hadrons in $e^+e^-$ Annihilation}
\footnote{Work supported in part by the Schweizerischer Nationalfonds}\\
\vskip 1cm
{\large Zoltan Kunszt and Zolt\'an  Tr\'ocs\'anyi}
\footnote{On leave from Kossuth University, Debrecen, Hungary}\\
\vskip 0.2cm
Theoretical Physics, ETH, \\
Zurich, Switzerland  \\
\vskip 1cm
\end{center}
\endgroup

\begin{abstract}
A detailed investigation of the theoretical ambiguities present in the
QCD description of  photon production in $e^+e^-$ annihilation is
given. It is pointed out that in a well-defined perturbative analysis
it is necessary to subtract the quark-photon collinear singularities.
This subtraction requires the introduction of an unphysical parameter
in the perturbative part of the cross section. The subtracted term is
factored into non-perturbative fragmentation function. The dependence
on the unphysical parameter cancels in the sum of non-perturbative
and perturbative parts. It is pointed out that for $E_{\gamma}\le
\sqrt{s}/(2(1+\epsilon_c))$ the non-perturbative contributions are
suppressed. Using a general purpose next-to-leading order Monte Carlo
program, we calculate various physical quantities that were measured
in LEP experiments recently.
\end{abstract}
\end{titlepage}
\setcounter{footnote}{0}

\section{Introduction}

The production of a photon (or an isolated photon) in association with
hadrons in $e^+e^-$ annihilation is a useful process to learn about the
differences in the properties of \q\qbar$\gamma$ and \q\qbar\g\ final
states, to measure the parton-photon fragmentation function and to test
QCD predictions in a channel crossed to photon-photon annihilation.  The
corresponding theoretical problems are well understood in the case of
prompt photon production at hadron colliders, photo-production of jets
and heavy flavor and photon-photon scattering.  It is an important
development that experiments at LEP give us high statistics data and
open ground to study even photon plus multijet final states
\cite{LEP,OPAL}. The better data call for a quantitative QCD
description.

The QCD description of inclusive photon production has a
simple, but important feature: the photon has hadronic component. In the
perturbative treament this fact is reflected by the appearance of
collinear photon-quark singularities. In order to obtain well defined
cross sections in \pQCD\ in all orders of the running coupling \als,
these singularities are to be subtracted and absorbed into the photon
fragmentation functions (factorization theorem) \cite{Aletal,FP}. The
fragmentation functions of the photon satisfy inhomogeneous evolution
equation; it grows with $Q^2$ therefore, it is called ``anomalous''
\cite{W,Detal,WZ}.

It is also interesting
to study the case of isolated photon. Physical isolation means that we
isolate the photons from hadrons and so we cannot make distinction
between quarks and gluons. Gluons, however, cannot be
isolated completely from the photon without destroying the cancellation
of soft gluon singularities between the virtual and real gluon
corrections.
Therefore, a physical isolation cannot eliminate
completely the collinear photon-quark singularities,
and so, even in the case of isolated photon production
the cross section contains ``anomalous'' (non-perturbative) piece.
This problem has been recognized clearly in the
next-to-leading order QCD study of isolated photon production at hadron
colliders \cite{ABF,BQ}.  The theoretical subtleties of defining
isolated photon cross section in \pQCD, however, have not been clearly
formulated in previous studies in the case of $e^+e^-$ annihilation
\cite{KL,KS}.

In section 2 we review the next-to-leading order description of
the inclusive (non-isolated) photon production.
In section 3 we outline the change in the formalism due to
the introduction of isolation cuts for the photon production.
We point out that isolation cannot completely eliminate
the non-perturbative fragmentation contribution, although it can
reduce its size. In section 4 a detailed perturbative study is
given for the cross section of  isolated photon plus jet
production up to order ${\cal O}(\alpha\alpha_s)$.
We review the mechanisms of the cancellation of the infrared
singularities and point out that in perturbation theory for processes
containing a photon in the final state the definition of
a  finite hard scattering cross requires a counter term which
necessarily  introduces an unphysical parameter.
Section 5 contains our numerical results for the isolated photon plus
$n$-jet production at LEP.  To demonstrate the flexibility of our
numerical program to calculate any jet shape parameters, we calculate
the distribution of the photon transverse momentum with respect to the
thrust axis as well.  The last section contains our conclusions.

\section{Inclusive photon production in $e^+ e^-$ annihilation}

According to the factorization theorem, the physical cross section
of inclusive photon production is obtained by folding
the fragmentation functions $D_{\gamma /a}(x,\mu_f)$ with the finite
hard-scattering cross sections $\d \hat{\sigma}_a$:
\begin{equation}
\label{inclusive}
\frac{\d \sigma_\gamma}{\d E_\gamma}=\sum_a
\int_0^{\sqrt{s}/2}\d E_a\,\int_0^1\d x\,D_{\gamma /a}(x,\mu_f)
\frac{\d \hat{\sigma}_a}{\d E_a}(E_a,\mu, \mu_f, \alpha_s(\mu))
\delta(E_\gamma- xE_a),
\end{equation}
where $\alpha_s(\mu)$ is the strong coupling constant at the ultraviolet
renormalization scale $\mu$ and $\mu_f$ is the factorization scale.

It is instructive to investigate the decomposition of this generally
valid expression up to next-to-leading order. First we remark that
\begin{equation}
\label{dff}
D_{\gamma/\gamma}(x)=\delta(1-x)+{\cal O}(\alpha^2),
\end{equation}
therefore, to leading order in the electromagnetic coupling, the term in
eq.\ (\ref{inclusive}) given by $a=\gamma$ is a purely perturbative
contribution. We use this equation to eliminate $D_{\gamma/\gamma}(x)$
from eq.\ (\ref{inclusive}). The  hard scattering cross section
$d\hat{\sigma}_{\gamma}/dE_{\gamma}$ is of order $\alpha$ in comparison
to the leading order annihilation cross section $\sigma_0$.\footnote{
In the following analysis, when the order of a contribution is given, it
is always understood in comparison to the leading order annihilation
cross section $\sigma_0$.} The leading non-perturbative part given by
the fragmentation function, however, is of order $\alpha/\alpha_s$.
This contribution is the ``anomalous'' photon component. Its
enhanced order is due to the fact that the scale dependence of the
fragmentation functions $D_{\gamma/a}(x,\mu_f)$, $a=q,\bar{q},g$ are
given by the inhomogeneous renormalization group equations \cite{KWZ,O}:
\begin{equation}
\mu\frac{\partial}{\partial\mu}D_{\gamma/a}(x,\mu)=
\frac{\alpha}{\pi} P_{\gamma/a}(x)
+ \frac{\alpha_s}{\pi} \sum_{b}
\int\frac{\d y}{y}\,D_{\gamma/b}\left(\frac{x}{y},\mu\right)P_{b/a}(y),
\end{equation}
where $P_{b/a}(x)$ denote the Altarelli-Parisi splitting functions.
To order $\alpha\alpha_s$ the inhomogeneous terms have the expressions
\cite{FP}
\begin{equation}
P_{\gamma/a}(x)=P^{(0)}_{\gamma/a}(x)+
\frac{\alpha_s}{2\pi}P^{(1)}_{\gamma/a}(x),
\end{equation}
where
\begin{equation}
\label{FPq}
P^{(0)}_{\gamma/q}(x)=e_q^2\frac{1+(1-x)^2}{x},
\quad\quad
P^{(0)}_{\gamma/g}(x)=0,
\end{equation}
and  after trivial replacement of  the color factors in eq.\ (12) of
ref.\ \cite{FP}, we have
\begin{eqnarray}
\lefteqn{P^{(1)}_{\gamma/q}(x)=} \\
& & e_q^2 C_F\left\{-\frac{1}{2}+\frac{9}{2}x+
\left(-8+\frac{1}{2}x\right)\log{x}+2x\log{(1-x)}+
\left(1-\frac{1}{2}x\right) \log^2{x}\right. \nonumber \\
& & \left.+\left[\log^2{(1-x)}+4\log{x}\log{(1-x)}+8\,{\rm Li}_2(1-x)-
\frac{4}{3}\pi^2\right]P^{(0)}_{\gamma/q(\bar{q})}(x)\right\},\nonumber
\end{eqnarray}
\begin{eqnarray}
\label{FPg}
P^{(1)}_{\gamma/g}(x)&=&\langle e_q^2\rangle T_R
\left\{-4+12x-\frac{164}{9}x^2+\frac{92}{9}x^{-1}\right.\nonumber \\
& &\left.+\left(10+14x+\frac{16}{3}x^2+\frac{16}{3}x^{-1}\right)\log{x}
+2(1+x)\log^2{x}\right\}. \nonumber
\end{eqnarray}
In the last equation,
\begin{equation}
\langle e_q^2 \rangle \equiv \sum_{q=1}^{N_F} e_q^2,
\end{equation}
where $N_F$ is the number of flavors.\footnote{We assume $e^+e^-$
annihilation via virtual photon. In order to obtain formulas valid at
the $Z^0$ peak, trivial modifications of charge factors are required.}

The unique solution of these inhomogeneous equations requires
non-perturbative input\footnote{In the literature it is usually called
Vector Meson Dominance (VMD) contribution \cite{GGReya,O,Aletal}.} at a
certain initial scale
$\mu$. At asymptotically large values of $\mu$, however, the solutions
are independent of the initial values and one obtains
\begin{eqnarray}
\label{asym1}
\lim_{\mu\rightarrow\infty}D_{\gamma/q}(x,\mu)&=&
\frac{\alpha}{2\pi}\log\frac{\mu^2}{\Lambda^2} a_{\gamma/q}(x),\\
\label{asym2}
\lim_{\mu\rightarrow\infty}D_{\gamma/g}(x,\mu)&=&
\frac{\alpha}{2\pi}\log\frac{\mu^2}{\Lambda^2} a_{\gamma/g}(x).
\end{eqnarray}
Exact analytic expressions for the Mellin transforms of the
$a_{a/\gamma}(x)$ functions have been found in refs.\ \cite{W,Detal}.
These are related to the $a_{\gamma/a}$ functions via crossing. It is
useful, however, to have a parametrization in $x$-space. Formulas which
accurately reproduce the exact leading logarithmic solutions were given
in ref.\ \cite{DukeO}:
\begin{eqnarray}
a_{\gamma/q}(x)&=&e_q^2\frac{1}{x}\left[
\frac{2.21-1.28x+1.29x^2}{1-1.63\log(1-x)}x^{0.049}
+0.002(1-x)^2x^{-1.54}\right],\\
a_{\gamma/g}(x)&=&\frac{1}{x}[0.0243(1-x)^{1.03}x^{-0.97}].
\end{eqnarray}
A new  parametrization of  the photon fragmentation functions
is described in ref.\ \cite{Auetal}.
The most striking feature of these solution is that they
increase as $1/\alpha_s$ with increasing the evolution scale.
Therefore, at high energy the contribution from the quark fragmentation
into a photon gives  the leading order ($\alpha/\alpha_s$) term
\begin{equation}
\frac{ \d\sigma_{\gamma}^{(0)} }{ \d E_{\gamma}}=
\sigma_0 \frac{4}{\sqrt{s}}\sum_q \frac{e_q^2}{\langle e_q^2\rangle}
D_{\gamma/q}\left(\frac{2E_\gamma}{\sqrt{s}},\mu\right)
+ {\cal O}(\alpha),
\end{equation}
In next-to-leading order, the $\mu$ dependence of $D_{\gamma/q}$ has to
be calculated with the next-to-leading order evolution equation and we
should also add the order $\alpha$ hard scattering cross section
\begin{eqnarray}
\lefteqn{\frac{\d\sigma_\gamma}{\d E_\gamma}=
\sigma_0\frac{4}{\sqrt{s}}\sum_q \frac{e_q^2}{\langle e_q^2\rangle}
D_{\gamma/q}\left(\frac{2E_{\gamma}}{\sqrt{s}},\mu\right)}\nonumber \\
&& + \sum_{a}\int_0^{\sqrt{s}/2}\d E_a\,\int_0^1\d x\,
D_{\gamma /a}(x,\mu_f)\frac{\d \hat{\sigma}^{(1)}_a}{\d E_a}
(E_a,\mu, \mu_f, \alpha_s(\mu))\delta(E_\gamma- xE_a) \nonumber  \\
&& + \ \frac{\d \hat{\sigma}^{(0)}_{\gamma}}{\d E_{\gamma}}
(E_{\gamma},\mu_f, \alpha_s(\mu))\ + \  {\cal O}(\alpha\alpha_s),
\end{eqnarray}
where $\d \hat{\sigma}^{(1)}_a/\d E_a$ denotes the order
$\alpha_s$ cross section of quark and gluon production.

The ${\cal O}(\alpha )$ hard-scattering cross section
$\d\hat{\sigma}_\gamma^{(0)}/\d E_\gamma$ is defined by
subtracting the photon-quark collinear singularity in the \ms\ scheme
\begin{equation}
\frac{\d\hat{\sigma}_\gamma^{(0)}}{\d E_\gamma}=
\lim_{\varepsilon\rightarrow 0}\left(\frac{\d\tilde{\sigma}^{(0)}}
{\d E_\gamma} +\frac{\d\sigma^{(0)}_{\rm CT}}{\d E_\gamma}\right),
\end{equation}
where the first term on the right hand side is the partonic cross
section in $4-2\varepsilon$ dimensions as defined by Feynman diagrams
\begin{equation}
\label{bare}
\frac{\d\tilde{\sigma}^{(0)}}{\d E_\gamma}=
\sigma_0\sum_q \frac{e_q^4}{\langle e_q^2 \rangle} \frac{\alpha}{2\pi}
\frac{2}{\sqrt{s}}H\left(\frac{4\pi\mu^2}{s}\right)^\varepsilon
\frac{1}{\Gamma(1-\varepsilon)}\int\d y_{12}\,\d y_{13}\,\d y_{23}\,
\theta(1- y_{13}-y_{23})(y_{12}y_{13}y_{23})^{-\varepsilon}
\end{equation}
\[ \times\left[(1-\varepsilon)\left(\frac{y_{23}}{y_{13}}
+\frac{y_{13}}{y_{23}}\right)
+\frac{2y_{12}-\varepsilon y_{13}y_{23}}{y_{13}y_{23}}\right]
\delta(1-y_{12}-y_{13}-y_{23})
\delta\left(1-y_{12}-\frac{2E_\gamma}{\sqrt{s}}\right),\]
where $H=1+{\cal O}(\varepsilon)$, while the second term
is the \ms\ counter-term
\begin{equation}
\label{CT}
\frac{\d\sigma^{(0)}_{\rm CT}}{\d E_\gamma}=\frac{\alpha}{2\pi}
\frac{(4\pi)^\varepsilon}{\varepsilon\Gamma(1-\varepsilon)}
\sum_q \int_0^{\sqrt{s}/2}\d E_q\,\int_0^1\d x\,
P^{(0)}_{\gamma/q}(x)\frac{\d\hat{\sigma}^{(0)}_q}{\d E_q}(E_q)
\delta(E_\gamma-xE_q).
\end{equation}
The integrations in eqs.\ (\ref{bare}), (\ref{CT}) are easily performed.
The collinear poles cancel in their sum. Setting $\varepsilon=0$,
one obtains
\begin{equation}
\label{hardf}
\frac{\d\hat{\sigma}_\gamma^{(0)}}{\d E_\gamma} =
\sigma_0\frac{\alpha}{2\pi}\frac{2}{\sqrt{s}}
\sum_{q=1}^{2N_F}
\frac{e^2_q}{\langle e_q^2\rangle}
P^{(0)}_{\gamma/q}(x_\gamma)\log\left(
\frac{s(1-x_\gamma)x^2_\gamma}{\mu^2}\right),
\end{equation}
where $x_\gamma=2E_\gamma/\sqrt{s}$.

The ${\cal O}(\alpha_s)$ corrections to the $\d\hat{\sigma}_{q,g}$
hard-scattering cross sections are defined by the Feynman diagrams
of fig.\ 1. First we note that $\d\hat{\sigma}^{(1)}_g$ can be
obtained from $\d\hat{\sigma}^{(0)}_\gamma$ by modifying the
charge factors:
\begin{equation}
\frac{\d\hat{\sigma}^{(1)}_g}{\d E_g}=C_F\sigma_0\frac{\alpha_s}{2\pi}
\frac{4}{\sqrt{s}}N_FP^{(0)}_{g/q}(x_g)
\log\left(\frac{s(1-x_g)x^2_g}{\mu^2}\right).
\end{equation}
The cross sections $\d\hat{\sigma}_q^{(1)}$ receives both real and
virtual corrections. The loop correction can be written as
\begin{equation}
\frac{\d\sigma_{\rm loop}}{\d E_q}=
C_F\sigma_0 \frac{\alpha_s}{2\pi}
\frac{2}{\sqrt{s}}H\left(\frac{\mu^2}{s}\right)^\varepsilon
\frac{(4\pi)^\varepsilon}{\varepsilon\Gamma(1-\varepsilon)}
\left[-\frac{2}{3}-3+(\pi^2-8)\varepsilon\right]
\delta\left(\frac{\sqrt{s}}{2}-E_q\right).
\end{equation}
The Bremsstrahlung contribution has an expression similar to
$\d\tilde{\sigma}^{(0)}/\d E_\gamma$ (\ref{bare}):
\begin{equation}
\frac{\d\sigma_{\rm real}}{\d E_q}=
C_F\sigma_0\frac{e_q^2}{\langle e_q^2\rangle}\frac{\alpha_s}{2\pi}
\frac{2}{\sqrt{s}}H\left(\frac{4\pi\mu^2}{s}\right)^\varepsilon
\frac{1}{\Gamma(1-\varepsilon)}\int\d y_{12}\,\d y_{13}\,\d y_{23}\,
\theta(1- y_{13}-y_{23})(y_{12}y_{13}y_{23})^{-\varepsilon}
\end{equation}
\[ \times\left[(1-\varepsilon)\left(\frac{y_{23}}{y_{13}}
+\frac{y_{13}}{y_{23}}\right)
+\frac{2y_{12}-\varepsilon y_{13}y_{23}}{y_{13}y_{23}}\right]
\delta(1-y_{12}-y_{13}-y_{23})
\delta\left(1-y_{23}-\frac{2E_\gamma}{\sqrt{s}}\right),\]
The sum of the loop and Bremsstrahlung contributions has the simple
expression
\begin{eqnarray}
\lefteqn{\frac{\d\tilde{\sigma}^{(1)}_q}{\d E_q}=
C_F\sigma_0 \frac{\alpha_s}{2\pi}\frac{2}{\sqrt{s}}H
\frac{(4\pi)^\varepsilon}{\varepsilon\Gamma(1-\varepsilon)}}\\
&\times &\nonumber \left\{-P^{(0)}_{q/q}(x_q)
+\varepsilon\left[P^{(0)}_{q/q}(x_q)\log\left(\frac{s}{\mu^2}\right)
+\left(\frac{2}{3}\pi^2-\frac{9}{2}\right)\delta(1-x_q)
+2\log x_q\frac{1+x_q^2}{1-x_q}\right.\right. \\
& &\nonumber \left.\left.
+(1+x_q^2)\left(\frac{\log(1- x_q)}{1-x_q}\right)_+
-\frac{3}{2}\left(\frac{1}{1-x_q}\right)_+-\frac{3}{2}x_q
+\frac{5}{2}\right]\right\},
\end{eqnarray}
where the index + denotes the usual ``+ prescription'' of regularizing
singular behavior at $x_q=1$. The remaining single pole is cancelled
when one adds the \ms\ counterterm $\d\sigma^{(0)}_{CT}$ which is
defined as
\begin{equation}
\frac{\d\sigma^{(1)}_{\rm CT}}{\d E_q}=
C_F\sigma_0\frac{\alpha_s}{2\pi}\frac{2}{\sqrt{s}}
\frac{(4\pi)^\varepsilon}{\varepsilon\Gamma(1-\varepsilon)}
P^{(0)}_{q/q}\left(\frac{2E_q}{\sqrt{s}}\right).
\end{equation}
The final result is obtained after setting $\varepsilon=0$:
\begin{eqnarray}
\lefteqn{\frac{\d\hat{\sigma}^{(1)}_q}{\d E_q}=
C_F\sigma_0 \frac{\alpha_s}{2\pi}\frac{2}{\sqrt{s}}}\\
&\times &\nonumber
\left\{P^{(0)}_{q/q}(x_q)\log\left(\frac{s}{\mu^2}\right)
+\left(\frac{2}{3}\pi^2-\frac{9}{2}\right)\delta(1-x_q)
+2\log x_q\frac{1+x_q^2}{1-x_q}\right. \\
& &\nonumber \left.
+(1+x_q^2)\left(\frac{\log(1- x_q)}{1-x_q}\right)_+
-\frac{3}{2}\left(\frac{1}{1-x_q}\right)_+-\frac{3}{2}x_q
+\frac{5}{2}\right\},
\end{eqnarray}
where $x_q=2E_q/\sqrt{s}$. This result can also be deduced after
replacing  trivial color factors from the coefficient functions of
inclusive single hadron production calculated in ref.\ \cite{Aletal}.

The theoretical input described in this section is sufficient to
extract the photon fragmentation functions from experimental data
in next-to-leading order accuracy. A complete analysis  requires
the measurement of the inclusive photon production cross section at
various energies.  The recent LEP data give information at the $Z$-pole.
Unfortunately, the data obtained at PETRA, LEP and TRISTAN suffer from
low statistics.  Needless to say that such an experimental study would
give very important complementary information on the fragmentation
functions of the photon obtained at hadron colliders.

\section{Inclusive isolated photon production in $e^+ e^-$ annihilation}

Let us now consider the inclusive photon cross section with photon
isolation. One can argue that due to isolation cuts the fragmentation
contribution is suppressed. As a consequence, isolation changes the
relative importance of the different contributions. It is reasonable to
consider the effect of isolation typically as an order $\alpha_s$ effect.
After imposing the isolation cuts, the fragmentation
contribution will be of order $\alpha$, i.\ e., the same order as the
order of the  pointlike perturbative cross section
$\d \hat{\sigma}^{(0)}_{\gamma}/\d E_{\gamma}$.
Isolation in practice can only be made with finite energy resolution.
Therefore, we require that in a cone of half angle $\delta_c$
around the photon three momentum the deposited energy be less than a
fraction $\epsilon_c$ of the photon energy. In experiments this
parameter $\epsilon_c$ has a value typically about 0.1.
Calculating $\d\hat{\sigma}^{(0)}_{\gamma,\:{\rm iso}}/\ds\d E_{\gamma}$,
we should insert a combination of $\theta$ functions in the phase space
integrals as follows
\begin{eqnarray}
\label{sfc}
S(\epsilon_c,\delta_c)&=&
\theta(\vartheta_{q\gamma}-\delta_c)
\theta(\vartheta_{\bar{q}\gamma}-\delta_c) \nonumber \\
&+&\theta(\vartheta_{q\gamma}-\delta_c)
\theta(\delta_c-\vartheta_{\bar{q}\gamma})
\theta(\epsilon_c E_{\gamma}-E_{\bar{q}}) \\ \nonumber
&+&\theta(\vartheta_{\bar{q}\gamma}-\delta_c)
\theta(\delta_c-\vartheta_{q\gamma})
\theta(\epsilon_c E_{\gamma}-E_q).
\end{eqnarray}
Let us require that
\[\epsilon_c < \frac{1}{2} \quad {\rm and }\quad
\sin^2\frac{\delta_c}{2} < \frac{1}{2} \]
and choose integration variables
\[ x_\gamma=\frac{2E_{\gamma}}{\sqrt{s}},\quad
y=\frac{y_{13}}{x_\gamma}.\]
We define the hard scattering cross section again with a collinear
counter-term
\begin{equation}
\frac{\d\hat{\sigma}_{\gamma,\:{\rm iso}}^{(0)}}{\d x_\gamma}=
\lim_{\varepsilon\rightarrow 0}
\left(\frac{\d\tilde{\sigma}^{(0)}_{\rm iso}}{\d x_\gamma}
+\frac{\d\sigma^{(0)}_{\rm CT,\:iso}}{\d x_\gamma}\right),
\end{equation}
where the first term in the right hand side is calculated as given by
Feynman diagrams in $4-2\varepsilon$ dimensions and for the
counter-term, we use the \ms-type expression
\begin{equation}
\label{CTiso}
\frac{\d\sigma^{(0)}_{\rm CT, iso}}{\d x_\gamma}=
2\sigma_0\frac{\alpha}{2\pi}\sum_q\frac{e^2_q}{\langle e_q^2\rangle}
\frac{(4\pi)^\varepsilon}{\varepsilon\Gamma(1-\varepsilon)}
P^{(0)}_{\gamma/q}(x_\gamma)\theta(x_\gamma-\frac{1}{1+\epsilon_c}).
\end{equation}
After performing the integration over $y$ and setting $\varepsilon=0$,
one obtains
\begin{equation}
\label{sig0iso}
\frac{\d\hat{\sigma}_{\gamma,\:{\rm iso}}^{(0)}}{\d x_\gamma}=
2\sigma_0\frac{\alpha}{2\pi}\sum_q \frac{e_q^2}{\langle e_q^2\rangle}
\left\{\left[P^{(0)}_{\gamma/q}(x_\gamma)
\log\frac{s(1-x_\gamma)x_\gamma^2y_m}{\mu^2(1-y_m)}
+ e_q^2x_\gamma(1-2y_m)\right]
\theta\left(x_\gamma-\frac{1}{1+\epsilon_c}\right)\right.
\end{equation}
\[
{}~~~~~+\left.P^{(0)}_{\gamma/q}(x_\gamma)\log\frac{1-y_c}{y_c}
-e_q^2x_\gamma(1-2y_c)\right\},
\]
where $y_c$ and $y_m$ are defined as follows
\[
y_c= \frac{1-x_\gamma}{1-x_\gamma \sin^2\frac{\delta_c}{2}}
\sin^2\frac{\delta_c}{2},\qquad
y_m={\rm min}\left\{y_c, 1+\epsilon_c - \frac{1}{x_\gamma}\right\}.
\]

One can make several comments on this result.
\begin{itemize}
\item The unisolated case can be recovered in the limit $\epsilon_c
\rightarrow \infty$ (cf.\ eq.\ (\ref{hardf})).
\item Imperfect isolation allows for a contribution from
the fragmentation: the photon looks isolated since the relatively soft
fragments surrounding it are not counted.
\item Assuming perfect energy resolution
($\epsilon_c=0$) we obtain vanishing counter term. In higher order,
however, we can not isolate the photon from the soft gluons completely
(we shall discuss this point in great detail in the next section),
therefore, one can not set the value of $\epsilon_c$ to zero.
\item In the leading logarithmic approximation one can define
a fragmentation function with isolation satisfying a modified
inhomogeneous evolution equation
\begin{eqnarray}
\lefteqn{\mu\frac{\partial}{\partial\mu}
D_{\gamma/a}(x,\mu,\epsilon_c)=} \\ \nonumber
&&\frac{\alpha}{\pi} P_{\gamma/a}(x)
\theta\left(x-\frac{1}{1+\epsilon_c}\right)
+ \frac{\alpha_s}{\pi} \sum_{b}\int\frac{\d y}{y}\,
D_{\gamma/b}\left(\frac{x}{y},\mu,\epsilon_c\right)P_{b/a}(y).
\end{eqnarray}
Clearly, if $D_{\gamma/a}(x,\mu)$ is a solution of the evolution
equation without isolation then
\begin{equation}
D_{\gamma/a}(x,\mu,\epsilon_c)=
D_{\gamma/a}(x,\mu)\theta\left(x-\frac{1}{1+\epsilon_c}\right)
\end{equation}
will be the solution of the evolution equation with isolation.  In
next-to-leading logarithmic approximation and/or choosing a different
counter-term (for example completely subtracting the contribution of
the singular region as defined by the third term of eq.\ (\ref{sfc})),
the isolated fragmentation can also be dependent on $\delta_c$
therefore, in general, one cannot simply identify the isolated fragmentation
with the non-isolated fragmentation in the high-$x$ region.
\end{itemize}
In next-to-leading order, the physical cross section of isolated photon
production is given by the terms as follows
\begin{eqnarray}
\label{isolated}
\lefteqn{\frac{\d\sigma_{\gamma}^{\rm iso}}
{\d E_{\gamma}}(\epsilon_c,\delta_c) =
\frac{\d \hat{\sigma}^{(0)}_{\rm iso}}{\d E_{\gamma}}
+ \frac{\d \hat{\sigma}^{(1)}_{\rm iso}}{\d E_{\gamma}}+
2\sigma_0 \frac{\alpha}{2\pi}
D_{\gamma/q}^{\rm iso}(\frac{2E_{\gamma}}{\sqrt{s}},\mu)
\theta(\frac{2E_{\gamma}}{\sqrt{s}} - \frac{1}{1+\epsilon_c})}
\nonumber \\
&+&\sum_a
\int_0^{\sqrt{s}/2}\d E_a\,\int_{\frac{1}{1+\epsilon_c}}^1
\d x\,D_{\gamma /a}^{\rm iso}(x,\mu_f)
\frac{\d \hat{\sigma}_a}{\d E_a}(E_a,\mu,\mu_f,\alpha_s(\mu))
\delta(E_\gamma- xE_a).
\end{eqnarray}
This decomposition is scheme dependent. The first term on the right hand
side of this equation has been calculated in the \ms\ scheme (see eq.\
(\ref{sig0iso}). It also appears useful to calculate the next-to-leading
order perturbative cross section $\d\hat{\sigma}^{(1)}/dE_{\gamma}$ in
the \ms\ scheme. This  requires the calculation of the next-to-leading
order splitting function $P_{\gamma/a}^{(1)}$ in the presence of
isolation cuts and a corresponding local subtraction term has to be found.
This is a complex but feasible calculation. Since such a result is not
yet available, in the next section we carry out the calculation of
$\d \hat{\sigma}^{(1)}_{\gamma}$ in a different subtraction scheme where
the photon is completely isolated from the quarks but not from soft
gluons (``cone subtraction''). In this scheme, in leading order, the
counter-term is vanishing and the cross section becomes independent of
$\epsilon_c$:
\begin{equation}
\label{sig0isocone}
\frac{\d\hat{\sigma}^{(0)}}
{\d E_{\gamma}} = 2\sigma_0\frac{\alpha}{2\pi}
\sum_q \frac{e_q^2}{\langle e_q^2\rangle}\left\{ P^{(0)}_{\gamma/q}
(x_{\gamma}) \ln\frac{1-y_c}{y_c} - x_{\gamma}(1-2y_c)\right\}.
\end{equation}
We note that the logarithmic divergence at $x_{\gamma}=1$ is the usual
soft singularity. Contrary to the case of the \ms\ scheme, with cone
subtraction the cross section is continuous and always positive (see
fig.\ 2 for comparison). One may argue that in this  scheme the
perturbative part is separated more efficiently, consequently  the
contributions of the non-perturbative terms (proportional to
$D^{\rm iso}_{\gamma/a}$) become relatively smaller.

In general we find that the non-perturbative terms contribute mainly
in the region $x_{\gamma} > 1/(1+\epsilon_c)$ thus we conclude that
the perturbative predictions appear to be reliable  for
$E_{\gamma} < \sqrt{s}/(2(1+\epsilon_c))$.

In the next section we present the results of our next-to-leading
order perturbative calculation of $\d\hat{\sigma}^{(1)}_{\rm iso}$
for isolated photon plus n-jet production. We conjecture
that a jet algorithm applied to the isolated photon hard scattering
cross section (eq.\ (\ref{isolated})) provides an infrared safe isolated
photon plus $n$-jet cross section. This is supported by the
fact that our isolation prescription does not influence the soft-gluon
structure of the cross section. If we can define a jet algorithm,
\begin{equation}
\frac{\d\hat{\sigma}_\gamma^{\rm iso}}{\d E_\gamma}(\delta_c,\epsilon_c)=
\frac{\d\hat{\sigma}^{\rm iso}_{\gamma+{\rm 1\:jet}}}{\d E_\gamma}
(\delta_c,\epsilon_c)+
\frac{\d\hat{\sigma}^{\rm iso}_{\gamma+{\rm 2\:jets}}}{\d E_\gamma}
(\delta_c,\epsilon_c)+
\frac{\d\hat{\sigma}^{\rm iso}_{\gamma+{\rm 3\:jets}}}{\d E_\gamma}
(\delta_c,\epsilon_c)+\ldots,
\end{equation}
such that every term on the right hand side is finite and we count every
particle only once, then isolated photon plus $n$-jet cross section
appears to be infrared safe.

We shall see  that the non-perturbative (``anomalous'') contributions
are important only in the case of photon plus 1 jet production when
the cross section is dominated by the $x_{\gamma}>1/(1+\epsilon_c)$
region.

\section{Isolated photon plus $n$-jet production}

In QCD, the differential cross section at ${\cal O}(\alpha_s^2)$ is
a sum of the real and virtual corrections:
\begin{equation}
\d\sigma=|M_4|^2 \d S^{(4)}+|M_3|^2 \d S^{(3)},
\end{equation}
where $\d S^{(n)}$ is the $n$-body phase space element with the flux
factor included.
For infrared safe quantities both terms on the right hand side are
separately divergent, but the sum is finite.  It is very difficult to
handle numerically this cancellation.  Fortunately, at least at one
loop, the divergencies can be cancelled analytically.  There are two
commonly used algorithms to achieve such a cancellation --- the
subtraction method \cite{ERT,KN} and the phase space slicing method
\cite{FSKS,GG}.  They both rely on the fact that after partial
decomposition $|M_4|^2$ can be written as a sum of terms with single pole
singularity.  Focusing our attention to the case of $q \bar{q} \gamma g$
final state, we find four such terms:
\begin{equation}
\label{partdec}
|M_4|^2=C_F\alpha\alpha_s\left(
\frac{M_{gq}}{y_{gq}}+\frac{M_{g\bar{q}}}{y_{g\bar{q}}}+
\frac{M_{\gamma q}}{y_{\gamma q}}
+\frac{M_{\gamma\bar{q}}}{y_{\gamma\bar{q}}}\right),
\end{equation}
where
\begin{equation}
y_{ij}=(p_i+p_j)^2/s,\qquad (s=M_Z^2)
\end{equation}
The pole part of each term is defined as
\begin{equation}
\frac{P_{ij}}{y_{ij}},\qquad {\rm where} \quad
P_{ij}=\lim_{y_{ij}\rightarrow 0}M_{ij}.
\end{equation}
It can be integrated analytically over either the whole or a part of the
phase space.  In this way, in general, we obtain analytical expressions
for the regularized divergencies of $\d\sigma^{(4)}$ which cancel
against the divergencies of the virtual corrections, $\d\sigma^{(3)}$
(KLN theorem).  When a photon in the final state is observed, the
cancellation mechanism described above does not apply to the $y_{\gamma
q(\bar{q})}$ poles. The reason for this is that the process is exclusive
in the photon and the virtual corrections  with the photon in the loop
cannot contribute for kinematical reasons.

To make the discussion more transparent, let us consider contributions
from the region where only $y_{qg}$ is small. The virtual corrections
can also be split into three terms
\begin{equation}
\label{m3}
|M_3|^2=C_F\alpha\alpha_s\left(
M_{gq}^{(3)}+M_{g\bar{q}}^{(3)}+M_f^{(3)}\right),
\end{equation}
such that $M_{gq}^{(3)}$ contains one half of the singularities, the
second term contains the other half and the third is the finite
part.\footnote{For the reader's convenience, we give the explicit
expressions for $M_{ij}$, $M_{kl}^{(3)}$ and $M_f^{(3)}$ in the
appendix.} (Notice that there are no $M_{\gamma q}^{(3)}$, $M_{\gamma
{\bar q}}^{(3)}$ terms.)  Then we shall concentrate on
\begin{equation}
\label{cs}
\frac{M_{ij}}{y_{ij}}\d S^{(4)}+M_{ij}^{(3)}\d S^{(3)}
\end{equation}
parts of the cross section.

In the subtraction method, one considers the combination
\begin{equation}
\label{subtraction}
\frac{M_{gq}}{y_{gq}}\d S^{(4)}-\frac{P_{gq}}{y_{gq}}
\d S^{(4\rightarrow 3)}+
\left(\int \frac{P_{gq}}{y_{gq}}\d S^{(g)}\right)\d S^{(3)}+
M_{gq}^{(3)}\d S^{(3)},
\end{equation}
where the integration over $\d S^{(g)}$ is meant to be an integral over
the gluon variables.  $\d S^{(4\rightarrow 3)}$ means the factorized
four-body phase space element in the limit when the gluon is soft or
collinear to the quark:  $\d S^{(4\rightarrow 3)}=\d S^{(g)}\d S^{(3)}$.

In the phase space slicing method, formula (\ref{cs}) is written as
\begin{equation}
\frac{M_{gq}}{y_{gq}}\theta(y_{gq}-y_0)\d S^{(4)}+
\frac{M_{gq}}{y_{gq}}\theta(y_0-y_{gq})\d S^{(4)}+M_{gq}^{(3)}\d S^{(3)}.
\end{equation}
If \y0\ is chosen small enough ($y_0\leq 10^{-4}$), then
\begin{equation}
\label{slice}
\frac{M_{gq}}{y_{gq}}\theta(y_{gq}-y_0)\d S^{(4)}+
\frac{P_{gq}}{y_{gq}}\theta(y_0-y_{gq})\d S^{(4\rightarrow 3)}+
M_{gq}^{(3)}\d S^{(3)}
\end{equation}
is a good approximation.  The first and second terms depend on \y0\
strongly, but their sum is independent of this unphysical parameter.
The strong \y0\ dependence originates mainly from the slicing of the
soft gluon region.

If one wishes to calculate isolated photon production, one has to make
sure that the restriction of the phase space does not disturb the
cancellation mechanism of soft and collinear gluons.  At hadron level
the meaning of photon isolation is well-defined.  At parton level
however, one has to be careful because the isolation prescription is
different for states with different number of partons.\footnote{One
may object isolation at parton level arguing that the fragmentation
process inevitably scatters hadronic matter into the isolation cone.
For a purely perturbative analysis, this objection is not valid. To
understand the reason for this, let us consider the same process at
higher energies, say $\sqrt{s}=10\,TeV$, in which energy region
perturbation theory is expected to give even better description.
Clearly, at this energy, fragmentation does not alter the energy flow,
therefore isolation at parton level corresponds to isolation at hadron
level.} If the photon is isolated form the partons with $y_\gamma$, we
should include isolation cuts with respect to all partons:
\begin{equation}
\theta(y_{\gamma q}-y_\gamma)\theta(y_{\gamma \bar{q}}-y_\gamma)
\theta(y_{\gamma g}-y_\gamma).
\end{equation}
The isolation from the gluon can be implemented only in the first term
of formula (\ref{subtraction}).  However, if we cut the soft gluons in
the first term, then the cancellation of singularities between the first
and second terms breaks down.  One can maintain the cancellation
introducing an energy resolution parameter $\epsilon$ such that a
gluon is isolated from the photon only if its energy is greater then
$\epsilon E_\gamma$.  Accordingly, isolation for the first term means
multiplication with
\begin{equation}
\label{iso}
\theta(y_{\gamma q}-y_\gamma)\theta(y_{\gamma \bar{q}}-y_\gamma)
(1-\theta(y_\gamma-y_{\gamma g})\theta(E_g-\epsilon E_\gamma)).
\end{equation}
Clearly, this criterium is ``not physical'' in the sense that one cannot
implement it at hadron level since we apply different cuts to quarks and
gluons.

If we introduce photon isolation in the slicing method, from formula
(\ref{slice}) we obtain
\begin{equation}
\label{isoslice}
\frac{M_{gq}}{y_{gq}}\theta(y_{gq}-y_0)\d S^{(4)}
\theta(y_{\gamma q}-y_\gamma)\theta(y_{\gamma \bar{q}}-y_\gamma)
\theta(y_{\gamma g}-y_\gamma)+
\end{equation}
\[
\left(\frac{P_{gq}}{y_{gq}}\theta(y_0-y_{gq})\d S^{(4\rightarrow 3)}+
M_{gq}^{(3)}\d S^{(3)}\right)
\theta(y_{\gamma q}-y_\gamma)\theta(y_{\gamma \bar{q}}-y_\gamma).
\]
Usually, $y_\gamma\gg y_0$.  This means that when changing \y0\ at fixed
\yg , the contribution from the soft gluons will be cut independently of
\y0\ and consequently, the \y0\ dependence is damped in the first term.
On the other hand, in the second term the \y0\ dependence is not damped
by the gluon-photon isolation condition.  The conclusion is that the
\y0\ dependence will be different in the two terms and, therefore, the
cross section will depend on the unphysical parameter \y0. It is
important to notice that if $y_{gq}>y_0$, then there exist an
$\epsilon'$ such that if $\epsilon<\epsilon'$ then
\begin{equation}
\label{equivalence}
1-\theta(y_\gamma-y_{\gamma g})\theta(E_g-\epsilon E_\gamma)=
\theta(y_{\gamma g}-y_\gamma),
\end{equation}
therefore (\ref{isoslice}) defines a finite cross section, but \y0\
plays in a sense the role of $\epsilon$ used in formula (\ref{iso}).

To demonstrate the \y0\ dependence of the isolated photon cross section
explicitly, we calculated the isolated photon plus 1- and 2-jet cross
sections using the isolation criterium given by formula
(\ref{isoslice}). First we make two technical remarks about the slicing
method.

Choosing large \y0, the pole approximation is not precise enough in the
singular region; one has to take into account the non-singular terms in
the same region, i.\ e., one should add the
\begin{equation}
\label{correction}
\left(\frac{M_{gq}}{y_{gq}}\d S^{(4)}-
\frac{P_{gq}}{y_{gq}}\d S^{(4\rightarrow 3)}\right)\theta(y_0-y_{gq})
\end{equation}
correction term.
The calculation becomes analogous to the subtraction method and one has
to introduce the $\epsilon$ energy resolution parameter.

It is a practical question to establish at what value of \y0\ the
correction term (\ref{correction}) becomes important.  The most
straightforward way to calculate the finite terms in (\ref{isoslice}) is
to perform the integration by a Monte Carlo method, which leaves
sufficient flexibility to calculate any jet shape parameter one wishes
to obtain.  The Monte Carlo calculation has a finite statistical error
which of course, can be reduced by generating more points.  Then the
criterium which determines the importance of the correction term
(\ref{correction}) is to require that the systematic error introduced by
neglecting (\ref{correction}) has to be smaller then the statistical
one.  Clearly, this critical value of \y0\ depends on the jet resolution
parameter \yj\ as well as on \yg.  For the case of 3-jet production
without photon in the final state, an analysis was carried out in
ref.\ \cite{GG} to determine the critical
value of \y0\ above which the systematic error dominates.  They found
that choosing $y_0/y_J\leq 0.01$ removes the systematic error.

The number of isolated photon plus $n$-jet events can be conveniently
parametrized in the form
\begin{eqnarray}
\frac{1}{\sigma_0}\sigma_{\gamma+n\:{\rm jets}}(y_J,y_0)&=&
\frac{\alpha}{2\pi}\sum_q\frac{e_q^4}{\langle e_q^2\rangle}
\left(g_n^{(0)}(y_J,y_0)
+\frac{\alpha_s}{2\pi}g_n^{(1)}(y_J,y_0)\right) \\ \nonumber
&\equiv&\frac{\alpha}{2\pi}\sum_q\frac{e_q^4}{\langle e_q^2\rangle}
g_n^{(0)}(y_J,y_0)(1+\alpha_s R_n(y_J,y_0)),
\end{eqnarray}
where $\sigma_0$ is the leading order cross section of the reaction
$e^+e^-\rightarrow$ hadrons and the $R_n(y_J,y_0)$ functions are defined
by the equation.  In figs.\ 3 and 4, we show the \y0\ dependence of the
${\cal O}(\alpha_s)$ QCD corrections, $R_n(y_J,y_0)$, to the isolated
photon plus 1-jet and the isolated photon plus 2-jet cross sections.  To
obtain the corrections, we used the following algorithm:
\begin{enumerate}
\item select isolated $\gamma+n$-jet events by requiring the invariant
mass of the photon with {\em any particle} in the event to be larger
than \yg\ (see formula (\ref{isoslice});
\item apply E0 cluster algorithm to the hadronic part of the event;
\item separate $\gamma+$ 1-, 2-, and 3-jet events by the number of
remaining clusters of hadrons.
\end{enumerate}
We used $y_\gamma=y_J$.  As expected, the \y0\ dependence in
$R_n(y_J,y_0)$ is strong up to $y_0=y_\gamma$.  As explained before, in
formula (\ref{isoslice}) the \y0\ cut plays the role of the
$\epsilon$ parameter of formula (\ref{iso}).  Therefore, the
(apparently) physical cut, (\ref{isoslice}) is in fact unphysical because
\y0\ is no longer a dummy variable of the cross section.

In order to demonstrate that we control the numerical evaluation of the
integrals at small \y0\ values, we calculated $R_n(y_J,y_0)$ with
$\theta(y_{\gamma g}-y_\gamma)$ in (\ref{isoslice}) removed.  We denote
the corresponding quantity with $\tilde{R}_n(y_J,y_0)$.  According to
the discussion after formula (\ref{isoslice}), this alteration should
remove the \y0\ dependence.  The explicit calculation shows that this
indeed happens.\footnote{In fact, we can see weak \y0\ dependence in
$\tilde{R}_n(y_J,y_0)$.  The origin of this dependence is the use of the
pole approximation.  It can be observed for values $y_0>10^{-3}$ (this
value depends on \yj ) in accordance with the observation made in ref.\
\cite{GG}. The results are shown for $y_J=0.06$. For other values of
\yj\ the dependence is similar.} We see that in order to obtain a \y0\
independent result, we have to use an unphysical cut:  different cuts
are applied to quarks and gluons.

\bigskip

We conclude from this discussion that if we want to define a finite
isolated photon plus $n$-jet cross section we have to make a subtraction
which depends on some unphysical parameter no matter which algorithm ---
the subtraction or slicing one --- is used (see formulas (\ref{iso}),
(\ref{isoslice}) and (\ref{equivalence})).  In other words, the physical
isolated photon plus $n$-jet cross section always contains some
non-perturbative (``anomalous'') contribution which is expected to give
contributions comparable or somewhat smaller than the ${\cal
O}(\alpha_s)$ QCD corrections.  For the separation of perturbative and
non-perturbative parts of the cross section, one must introduce an
unphysical (non-zero) parameter.  Of course, the sum of the perturbative
and non-perturbative pieces is independent of this parameter. In the
previous section we pointed out that the non-perturbative contribution
is expected to be small for $x_\gamma<1/(1+\epsilon_c)$.

\section{Numerical results}

As advocated in section 4, we carry out our calculation with the
subtraction method. The event definition is the following:
\begin{enumerate}
\item isolate the photon;
\item apply E0 cluster algorithm to the hadronic part of the event;
\item separate $\gamma+$ 1-, 2-, and 3-jet events by the number of
remaining clusters of hadrons.
\end{enumerate}
Photon isolation can be achieved either by isolating the photon in a
cone (cone isolation) or by requiring the invariant mass of the photon
with any particle in the event to be larger then an invariant mass cut
\yg . From experimental point of view the cone isolation is more natural.
Unfortunately, the results by OPAL \cite{OPAL} are corrected experimental
values in order to compare the measured rates with the matrix element
calculation of \cite{KL} where invariant mass isolation was used (with
$y_\gamma=y_J$). We give results for both. Since the QCD corrections are
very sensitive to the event definition  we give explicitly how formula
(\ref{subtraction}) is modified in the case of cone isolation:
\begin{eqnarray}
\label{ci}
\lefteqn{
\theta(\vartheta_{q\gamma}-\delta_c)
\theta(\vartheta_{\bar{q}\gamma}-\delta_c)} \nonumber \\
&\times& \left\{
(1-\theta(\delta_c-\vartheta_{g\gamma})
\theta(E_g-\epsilon_c E_\gamma))
\frac{M_{gq}}{y_{gq}}\d S^{(4)}\right. \\ \nonumber
& &\left. -\frac{P_{gq}}{y_{gq}} \d S^{(4\rightarrow 3)}+
\left(\int \frac{P_{gq}}{y_{gq}}\d S^{(g)}\right)\d S^{(3)}
+M_{gq}^{(3)}\right\};
\end{eqnarray}
and in the case of invariant mass isolation:
\begin{eqnarray}
\label{imi}
\lefteqn{
\theta(y_{q\gamma}-y_J) \theta(y_{\bar{q}\gamma}-y_J)}\nonumber \\
&\times& \left\{
(1-\theta(y_J-y_{g\gamma})\theta(E_g-\epsilon_c E_\gamma))
\frac{M_{gq}}{y_{gq}}\d S^{(4)}\right. \\ \nonumber
& &\left.-\frac{P_{gq}}{y_{gq}} \d S^{(4\rightarrow 3)}+
\left(\int \frac{P_{gq}}{y_{gq}}\d S^{(g)}\right)\d S^{(3)}
+M_{gq}^{(3)}\right\}.
\end{eqnarray}

To obtain the isolated photon plus $n$-jet rates, the formulas above are
multiplied with $\theta$ functions as follows:
\begin{itemize}
\item One photon plus 3-jet:
\begin{equation}
\theta(y_{qg}-y_J) \theta(y_{\bar{q}g}-y_J) \theta(y_{q\bar{q}}-y_J)
\theta(y_{q\gamma}-y_J) \theta(y_{\bar{q}\gamma}-y_J)
\theta(y_{g\gamma}-y_J).
\end{equation}
\item One photon plus 2-jet:\\
Denote $i$ and $j$ the partons which when combined have the smallest
invariant mass in the hadronic part of the event, so they are combined
into pseudoparticle $c$. Denote $k$ the third parton. In the three-body
phase space the momentum of the $j$ particle is identically zero. Then
we use
\begin{equation}
\theta(y_J-y_{ij})\theta(y_{ck}-y_J)
\theta(y_{c\gamma}-y_J) \theta(y_{k\gamma}-y_J).
\end{equation}
\item One photon plus 1-jet:
\begin{equation}
\label{1g1j}
\theta(y_J-y_{qg}) \theta(y_J-y_{\bar{q}g})
\theta(y_J-y_{q\bar{q}}) \theta(y_J-y_{ck}).
\end{equation}
\end{itemize}
In the case of cone isolation, we also required that the energy of the
photon has to be larger than 7.5\,GeV. The half-opening angle of the
cone is $15^\circ$.

We shall give the results of our calculation for the partial widths
$\Gamma(Z\rightarrow\gamma+n\:{\rm jets})$ as ratios to the hadronic
width:
\begin{equation}
\frac{\Gamma(Z\rightarrow\gamma+n\:{\rm jets})}
{\Gamma(Z\rightarrow{\rm hadrons})}=
\frac{\left(\frac{\ds 8}{\ds 9}c_u+\frac{\ds 1}{\ds 3}c_d\right)
\frac{\ds \alpha}{\ds 2\pi}}{(2c_u+3c_d)
\left(1+\frac{\ds \alpha_s}{\ds \pi}
+1.42\left(\frac{\ds \alpha_s}{\ds \pi}\right)^2\right)}g_n(y_J),
\end{equation}
where
\begin{equation}
c_f=v_f^2+a_f^2
\end{equation}
and $v_f$ and $a_f$ are the weak vector and axial vector couplings:
\begin{equation}
v_f=2I_{3,f}-4e_f\sin^2\theta_W
\end{equation}
\begin{equation}
a_f=2 I_{3,f},
\end{equation}
so with $\sin^2\theta_W=0.23$, $v_u=0.39$, $v_d=-0.69$, $a_u=+1$ and
$a_d=-1$. The $g_n(y_J)$ functions can be expanded in \als :
\begin{equation}
g_n(y_J)=g_n^{(0)}(y_J)+\frac{\alpha_s}{2\pi}g_n^{(0)}(y_J)
\equiv g_n^{(0)}(y_J)(1+\alpha_s R_n(y_J)),
\end{equation}
and our aim is to compute the $g_n^{(i)}$ functions (of course,
$g_3^{(0)}(y_J)=0$.)

It is interesting to study the photon energy spectrum of the jet cross
sections
\begin{equation}
\frac{\d \sigma^{\rm iso}_{\gamma + 1jet}}{\d E_{\gamma}}
(\delta_c,\epsilon_c), \quad
\frac{\d \sigma^{\rm iso}_{\gamma + 2jet}}{\d E_{\gamma}}
(\delta_c,\epsilon_c)
\end{equation}
at some realistic values of the isolation parameters $\delta_c,
\epsilon_c$. In fig.\ 5 the photon energy distibutions of one jet
and two jet production are shown in the Born approximation, while  in
fig.\ 6 the same curves are plotted but including the next-to-leading
order corrections. Due to obvious kinematical reasons one jet
production is completely dominated by the hard photon region
$x_{\gamma}>1/(1 + \epsilon_c)$. In this region as we pointed out
one may get substantial (but not overwhelmingly large) ``anomalous''
photon contribution. It is difficult to estimate the ``anomalous''
contribution since we do not have yet enough phenomenological input.
Certainly a combined study of the hadron collider and LEP data would
help to understand its size better. We note that in the high $x$ region
the application of perturbative QCD by itself requires some care due
to the appearance of large logarithms of type $\log(1-x)$. Indeed the
QCD corrections are larger for one jet than for two jet production.
It is interesting to compare the one jet data to the perturbative
QCD prediction, but one should not be surprised if one does not find
perfect agreement.
The non-perturbative corrections appear, however, negligible
in the case of 2-jet production since it is dominated by
the complementary region $x_{\gamma} < 1/(1+\epsilon_c)$. Requiring
that
$x_{\gamma} < 1/(1+\epsilon_c)$, the 2-jet results remain practically
unaffected, while this cut largely eliminates  the 1-jet production.
This is illustrated by the numbers given in Table 1.
There is a tendency that if we shrink the isolation region
the perturbative contribution increases.
{}From figs.\ 5, 6 and, we can see also that
the total one jet and two jet rates should depend
weekly on $\epsilon_c$. The reason for this is that
the photon energy distribution changes weakly
if we change $\epsilon_c$ in the physically interesting
region of 0.06--0.2 .

In addition to the ambiguities due to ``anomalous''
photon production there are also the usual scale ambiguities.
In fig.\ 7 we present the predicted values of the $\Gamma(Z\rightarrow
\gamma+n\:{\rm jets})$ ($n=1,2$) partial widths for the cone isolation
with $\epsilon_c=0.1$.  The bands between the dashed lines represent
the scale dependence between the scales $M_Z/2$ and
$2M_Z$.  We used $\alpha_s(M_Z)=0.12$ and $\alpha=1/137$. The
$\epsilon_c$ dependence is so weak for experimentally feasible values
that the uncertainty introduced by the $\epsilon_c$ dependendce is
much smaller than the scale dependence and therefore we did not show it.
The scale dependence of the 1-jet rate is rather large. This is a
reflection of the fact that the QCD corrections are large.
In figs.\ 8 and 9 the same curves  as in
fig.\ 7 are depicted in the case of invariant mass isolation
with $y_{\gamma}=y_J$ for the 1-jet and 2-jet rates, respectively.
In the same figures, we show the enhancement induced by
the choice of smaller isolation region. In accordance with our previous
discussion, the enhancement is larger for the 1-jet rate than for the
2-jet rate. We note, however, that when comparison is made to the data
at a given isolation it is important to use exactly the same
isolation and event definition both in the experimental
and theoretical analysis. Therefore one can not just change
the value of $y_{\gamma}$ such that the prediction fits
better  the data. In particular one is not
allowed to use different values of $y_{\gamma}$ in case of one jet
and two jet production. In a given   subtraction scheme with
well defined  experimental isolation cuts all the parameters of the
perturbative part are fixed.
In particular the discrepancy between the measured
$\gamma$+ 1-jet rate and the perturbative prediction at
$y_{\gamma}=y_J$ may indicate  non-negligible
anomalous contributions.

As mentioned in the section 4, the Monte Carlo approach is useful
because it leaves sufficient flexibility to calculate any jet shape
parameter.  To demonstrate this feature of our work we present the
result of matrix element calculation for the distribution of the photon
transverse momentum with respect to the thrust axis (fig.\ 10).  The
thrust axis has been calculated all particles taken into account,
including the photon.  We used invariant mass isolation (with
$y_\gamma=0.005$ and 0.06) to isolate the photon from the partons.  We
also required the photon to be more energetic than 7.5\,GeV. For
small $p_T$, configurations with thrust value close to one may occur.
 The histogram is normalized to one, therefore the
uncertainty in the small $p_T$ region influences the behaviour in the
large $p_T$ region. We note, however, that requiring
$x_{\gamma} < 1/(1+\epsilon_c)$ the small $p_T$ region will be
suppressed.

Finally, in fig.\ 11, we present the predicted values of the
$\Gamma(Z\rightarrow \gamma+n\:{\rm jets})$ ($n=1,2$) partial widths for
the cone isolation with $\epsilon_c=0.1$ when Durham clustering
algorithm is used \cite{D}. In this algorithm, two jets are combined in
to a single jet if
\begin{equation}
y_{Dij}=\frac{2{\rm min}(E_i^2,E_j^2)(1-\cos\theta_{ij})}{s}
\end{equation}
is smaller than the jet resolution parameter \yj. For pure QCD events,
this algorithm tends to emphasize 2-jet events as compared to other
algorithms and suited better for resummation purposes \cite{CDW}. When a
photon in the final state is observed, we find higher 1-jet rate and
lower 2-jet rate and the QCD corrections are smaller as compared to the
E0 cluster algorithm.

\section{Conclusions}

Photon production in association with hadrons in $e^+e^-$ annihilation
provides us interesting information on the non-perturbative
component of the photon and new possibilities to test the underlying
structure of perturbative QCD.

In this paper we paid special attention to the importance of the correct
treatment of the collinear photon-quark region.  It was shown that
next-to-leading and higher orders the perturbative part can only be
defined using some non-physical parameter, no matter whether
non-isolated or isolated photon production is considered.  The physical
cross section defined as the sum of the perturbative and non-perturbative
part is, of course, independent of such a parameter.

We briefly reviewed the theoretical description of the inclusive
non-isolated photon production in $e^+e^-$ annihilation.  It was pointed
out that the LEP data can be used to constrain the parametrization of
the fragmentation functions of the photon, $D_{\gamma/q}(x,\mu)$,
$D_{\gamma,g}(x,\mu)$.  The measurement of these fragmentation functions
would give important input information for the other inclusive photon
production measurements at hadron colliders and at HERA.  Furthermore,
one could test the anomalous $\mu$-dependence at asymptotically large
$\mu$ values predicted by perturbative QCD.

The case of isolated photon production was studied as well.  Under well
defined circumstances, isolation can suppress the numerical contribution
of the non-perturbative contributions.  We pointed out that the
non-perturbative (``anomalous'') contribution can be sizable only for
$E_{\gamma} > \sqrt{s}/2/(1+\epsilon_c)$, where $\epsilon_c$ is the
energy fraction in the isolation cone with respect to the photon energy.
When a jet algorithm is used, then the non-perturbative contribution is
expected to be further suppressed for isolated photon plus $n$-jet
cross section for $n>1$, but not for $n=1$.

We demonstrated the difficulty due to the quark photon collinear
singularity with careful calculation of the next-to-leading order QCD
corrections to isolated photon plus one or two jets. We argued
that in the case of isolated photon plus 2-jet
production indeed, as suggested by Kramer and Lampe, the
perturbative contribution dominates the physical cross section.
The next-to-leading order corrections are calculated
by developing  a Monte Carlo program
which can be used to calculate the perturbative corrections
to any physical quantity.

\bigskip\bigskip
\noindent {\bf \large Acknowledgement}\ \ We thank P. M\"attig and
C. Markus for helpful correspondence. One of us (Z.K.) is greatful to
R. K. Ellis and G. Sterman for illuminating discussions.

\bigskip\bigskip
\Large
{\bf Appendix A}
\normalsize

\bigskip

In this appendix we give the explicit expressions for the $M_{ij}$ and
$M^{(3)}_{ij}$ expressions used in formulas (\ref{partdec}) and
(\ref{m3}) respectively. We shall make the following renaming:
\begin{eqnarray}
&q& \rightarrow {\rm particle}\;1 \nonumber\\
&\bar{q}& \rightarrow {\rm particle}\;2 \\ \nonumber
&g& \rightarrow {\rm particle}\;3 \\ \nonumber
&\gamma& \rightarrow {\rm particle}\;4.
\end{eqnarray}
Then the following relations are valid:
\begin{equation}
M_{23}=M_{13}(1\leftrightarrow 2),\qquad
M_{14}=M_{13}(3\leftrightarrow 4),\qquad
M_{24}=M_{13}(1\leftrightarrow 2,\quad 3\leftrightarrow 4),
\end{equation}
therefore, it is sufficient to spell out $M_{13}$. The corresponding
expression can be obtained from the four-parton matrix element given in
Appendix B of ref.\ \cite{ERT} by setting $N_C=0$ and $T_R=0$. After
performing partial fractioning one obtains
\begin{eqnarray*}
M_{13}&=&\frac{2}{4\pi^2}\left[
\frac{2y_{12} y_{123} (1+y_{34})}{y_{134}y_{234}(y_{13}+y_{23})}
+\frac{2y_{14}(1-y_{24})}{y_{234} (y_{13}+y_{24})}
+\frac{2(1-y_{13}) y_{23}}{y_{134} (y_{13}+y_{24})}\right. \\
& &+\frac{1}{y_{134}^2}
(y_{24} y_{34}+y_{12} y_{34}+y_{13} y_{24}-y_{14} y_{23}+y_{12} y_{13})
+\frac{y_{34}}{y_{13}+y_{24}} \\
& & +\frac{y_{12} y_{24} y_{34}+y_{12} y_{14} y_{34}-y_{13}
y_{24}^2+y_{13} y_{14} y_{24}+2y_{12} y_{14} y_{24}}
{y_{134}(y_{13}+y_{14}+y_{23})}\left(
\frac{1}{y_{13}+y_{23}}+\frac{1}{y_{13}+y_{14}}\right) \\
& &+\frac{y_{12} y_{24} y_{34}+y_{12} y_{14} y_{34}-y_{14}^2
y_{23}+y_{14} y_{23} y_{24}+2y_{12} y_{14} y_{24}}
{y_{234}(y_{13}+y_{23}+y_{24})}\left(
\frac{1}{y_{13}+y_{23}}+\frac{1}{y_{13}+y_{24}}\right) \\
& &+\frac{y_{12} y_{23} y_{34}+y_{12} y_{13} y_{34}-y_{14}
y_{23}^2+y_{13} y_{14} y_{23}+2y_{12} y_{13} y_{23}}
{y_{134}(y_{13}+y_{14}+y_{24})}\left(
\frac{1}{y_{13}+y_{14}}+\frac{1}{y_{13}+y_{24}}\right)
\end{eqnarray*}
\begin{eqnarray*}
{}~~~~~~~+\frac{2y_{12} y_{123} y_{124}}{y_{13}+y_{23}+y_{14}+y_{24}}
& &\left(\frac{1}{y_{13}+y_{23}+y_{24}}
\left(\frac{1}{y_{13}+y_{24}}+\frac{1}{y_{13}+y_{23}}\right)\right.\\
& &+\frac{1}{y_{13}+y_{14}+y_{24}}
\left(\frac{1}{y_{13}+y_{14}}+\frac{1}{y_{13}+y_{24}}\right)\\
& &+\left.\frac{1}{y_{13}+y_{14}+y_{23}}
\left(\frac{1}{y_{13}+y_{14}}+\frac{1}{y_{13}+y_{23}}\right)\right) \\
+\frac{1}{y_{134} y_{234} (y_{13}+y_{24})}& &
(y_{12} y_{34}^2-y_{13} y_{24} y_{34}+y_{14} y_{23} y_{34}+3y_{12}
y_{23} y_{34} \\
& &+3y_{12} y_{14} y_{34}+4y_{12}^2 y_{34}
-y_{13} y_{23}y_{24}+2y_{12} y_{23} y_{24} \\
& &-y_{13} y_{14} y_{24}-2y_{12} y_{13} y_{24}
+2y_{12}^2 y_{24}+y_{14} y_{23}^2 \\
& &+2y_{12} y_{23}^2
+y_{14}^2 y_{23}+4y_{12} y_{14} y_{23}+4y_{12}^2y_{23} \\
& &+2y_{12} y_{14}^2
+2y_{12} y_{13} y_{14}+4y_{12}^2y_{14}+2y_{12}^2y_{13}+2y_{12}^3) \\
-\frac{1}{y_{134}(y_{13}+y_{14})}
& &(y_{14} y_{24}+2y_{14} y_{23}+2y_{12} y_{14}+y_{13}^2 \\
& &+y_{13} y_{23}+2y_{13} y_{24}+2y_{12} y_{13}+y_{14}^2)
\left.\frac{~}{~}\right]
\end{eqnarray*}
Due to partial fractioning, this expression is finite if a single
$y_{ij}\rightarrow 0$ (and for the same reason the expression is
lengthy.)

The virtual corrections can also be obtained easily from eq. (2.20) of
ref.\ \cite{ERT} by setting $N_C=0$ and $T_R=0$. In our decomposition
\begin{equation}
M_{gq}^{(3)}=M_{g{\bar q}}^{(3)}=
\frac{1}{4\pi^2}\left[-\frac{1}{\varepsilon^2}-
\frac{1}{2\varepsilon}(3-2\log{y_{12}})\right],
\end{equation}
and the finite part is
\begin{eqnarray}
\label{m3finite}
M_f^{(3)}&=&\frac{1}{4\pi^2}\left[
\frac{y_{12}}{y_{12}+y_{14}}+\frac{y_{12}}{y_{12}+y_{24}}+\right.
\frac{y_{12}+y_{24}}{y_{14}}+\frac{y_{12}+y_{14}}{y_{24}} \\ \nonumber
& &+\log{y_{14}}
\left[\frac{4y_{12}^2+2y_{12}y_{14}+4y_{12}y_{24}+y_{14}y_{24}}
{(y_{12}+y_{24})^2}\right] \\ \nonumber
& &+\log{y_{24}}
\left[\frac{4y_{12}^2+2y_{12}y_{24}+4y_{12}y_{14}+y_{14}y_{24}}
{(y_{12}+y_{14})^2}\right] \\ \nonumber
& &-2\left[
\frac{y_{12}^2+(y_{12}+y_{14})^2}{y_{14}y_{24}}R(y_{12},y_{24})
+\frac{y_{12}^2+(y_{12}+y_{24})^2}{y_{14}y_{24}}R(y_{12},y_{14})\right.\\
\nonumber & &~~~~~+\frac{y_{14}^2+y_{24}^2}{y_{14}y_{24}(y_{14}+y_{24})}-
2\log{y_{12}}\left.\left(
\frac{y_{12}^2}{(y_{14}+y_{24})^2}+\frac{2y_{12}}{y_{14}+y_{24}}\right)
\right]\\ \nonumber
& &\left.+\left(\frac{y_{24}}{y_{14}}+\frac{y_{14}}{y_{24}}+
\frac{2y_{12}}{y_{14}y_{24}}\right)
\left(\frac{2}{3}\pi^2-\log^2y_{12}-8\right)\right],
\end{eqnarray}
where
\begin{eqnarray}
\lefteqn{R(x,y)=} \\ \nonumber
& &\log x\log y-\log x\log (1-y)-\log y\log (1-x)
+\frac{1}{6}\pi^2-{\rm Li}_2(x)-{\rm Li}_2(y)
\end{eqnarray}
and
\begin{equation}
{\rm Li}_2(x)=-\int_0^x\d z\,\frac{\log (1-z)}{z}.
\end{equation}

\newpage

\Large
{\bf Figure captions}
\normalsize

\bigskip

\begin{description}
\item[Figure 1] Typical Feynman diagrams contributing to the
calculation of the ${\cal O}(\alpha_s)$ corrections to the inclusive
quark production in $e^+e^-$ annihilation.
\item[Figure 2] Leading order hard-scattering cross section for
inclusive isolated photon production calculated in the \ms\
subtraction scheme (eq.\ \ref{sig0iso}) and in the cone subtraction
scheme (eq.\ \ref{sig0isocone}). $\delta_c=15^\circ$ and
$\epsilon_c=0.1$ isolation parameters were used.
\item[Figure 3] The dependence of the QCD corrections to the isolated
photon plus 1-jet production on the unphysical parameter \y0\ when
physical cuts are applied (see formula (\ref{isoslice})) --- solid
curves --- and with unphysical cuts (only quarks are cut) --- dashed
curves. The slicing method in the pole approximation was used with
$y_\gamma=y_J$.
\item[Figure 4] The dependence of the QCD corrections to the isolated
photon plus 2-jet production on the unphysical parameter \y0\ when
physical cuts are applied (see formula (\ref{isoslice})) --- solid
curves --- and with unphysical cuts (only quarks are cut) --- dashed
curves. The slicing method in the pole approximation was used with
$y_\gamma=y_J$.
\item[Figure 5] Leading order photon energy spectrum of the partial
widths $\Gamma(Z\rightarrow\gamma+n\:{\rm jets})$, ($n=1,2$)
normalized to $10^{-3} \Gamma(Z \rightarrow {\rm hadrons})$ at
$y_J=0.1$, $\delta_c=15^\circ$ and $\epsilon_c=0.1$.
\item[Figure 6] Next-to-leading order photon energy spectrum of the
partial widths $\Gamma(Z\rightarrow\gamma+n\:{\rm jets})$, ($n=1,2$)
normalized to $10^{-3} \Gamma(Z \rightarrow {\rm hadrons})$ at
$y_J=0.1$, $\delta_c=15^\circ$ and $\epsilon_c=0.1$.
\item[Figure 7] Partial widths $\Gamma(Z\rightarrow\gamma+n\:{\rm
jets})$, ($n=1,2$) as a function of \yj\ normalized to $10^{-3} \Gamma(Z
\rightarrow {\rm hadrons})$ when cone isolation of the photon is used
(solid lines) with $\alpha_s(M_Z)=0.12$, $\alpha=1/137$. The dashed
curves represent the scale dependence between scales $\mu=M_Z/2$ and
$\mu=2M_Z$.
\item[Figure 8] Partial width $\Gamma(Z\rightarrow\gamma+1\:{\rm
jet})$, as a function of \yj\ normalized to $10^{-3} \Gamma(Z
\rightarrow {\rm hadrons})$ (solid line) when invariant mass isolation
is used with $y_\gamma=y_J$, $\epsilon_c=0.1$, $\alpha_s(M_Z)=0.12$,
$\alpha=1/137$.  The dashed curves represent the scale dependence
between scales $\mu=M_Z/2$ and $\mu=2M_Z$. The dashed dotted
curve is the partial width calculated with $y_\gamma=0.005$ and the
long-dashed short-dashed curve is that with $y_\gamma=0.001$.
\item[Figure 9] Partial width $\Gamma(Z\rightarrow\gamma+2\:{\rm
jets})$, as a function of \yj\ normalized to $10^{-3} \Gamma(Z
\rightarrow {\rm hadrons})$ (solid line) when invariant mass isolation
is used with $y_\gamma=y_J$, $\epsilon_c=0.1$ ,$\alpha_s(M_Z)=0.12$,
$\alpha=1/137$.  The dashed curves represent the scale dependence
between scales $\mu=M_Z/2$ and $\mu=2M_Z$. The long-dashed short-dashed
curve is the partial width calculated with $y_\gamma=0.001$.
\item[Figure 10] Distribution of the photon transverse momentum with
respect to the thrust axis.  The photon was isolated using invariant
mass isolation with $y_\gamma=0.005$ and 0.06.  The dotted histograms
show the scale dependence between scales $\mu=M_Z/2$ and $\mu=2M_Z$.
The width of one bin is $M_z/100$.
\item[Figure 11] Partial widths $\Gamma(Z\rightarrow\gamma+n\:{\rm
jets})$, ($n=1,2$) as a function of $y_D$ normalized to $10^{-3}
\Gamma(Z \rightarrow {\rm hadrons})$ when Durham clustering algorithm
and cone isolation of the photon is used (solid lines) with
$\alpha_s(M_Z)=0.12$, $\alpha=1/137$.  The dashed curves represent the
scale dependence between scales $\mu=M_Z/2$ and $\mu=2M_Z$.
\end{description}
\end{document}